\documentclass[sigconf]{acmart}
\setcopyright{cc}
\acmConference[STAIG '25]{CHI'25 Sociotechnical AI Governance Workshop}{April 2025}{Yokohama, Japan}
\acmYear{2025}
\copyrightyear{2025}
\renewcommand\footnotetextcopyrightpermission[1]{}

\usepackage{graphicx}
\AtBeginDocument{%
  }

\setcopyright{acmlicensed}
\copyrightyear{2025}

\begin{document}

\title[Limitations in Technical Governance of AI-Generated Non-Consensual Intimate Images of Adults]{The Malicious Technical Ecosystem: Exposing Limitations in Technical Governance of AI-Generated Non-Consensual Intimate Images of Adults
}
\author{Michelle L. Ding}
\email{michelle_ding@brown.edu}
\orcid{0009-0000-9778-1306}
\affiliation{%
  \institution{Brown University}
  \city{Providence}
  \state{Rhode Island}
  \country{USA}
  }

\author{Harini Suresh}
\orcid{0000-0002-9769-4947}
\email{harini_suresh@brown.edu}
\affiliation{%
  \institution{Brown University}
  \city{Providence}
  \state{Rhode Island}
  \country{USA}
}


\begin{abstract}
In this paper, we adopt a survivor-centered approach to locate and dissect the role of sociotechnical AI governance in preventing AI-Generated Non-Consensual Intimate Images (AIG-NCII) of adults, colloquially known as ``deep fake pornography.'' We identify a ``malicious technical ecosystem'' or ``MTE,'' comprising of open-source face-swapping models and nearly 200 ``nudifying'' software programs that allow non-technical users to create AIG-NCII within minutes. Then, using the National Institute of Standards and Technology (NIST) AI 100-4 report as a reflection of current synthetic content governance methods, we show how the current landscape of practices fails to effectively regulate the MTE for adult AIG-NCII, as well as flawed assumptions explaining these gaps.
\end{abstract}

\maketitle

\begin{center}
\smallskip
\begin{minipage}{\columnwidth}
\small
\textbf{ACM Reference Format:}\\
Michelle L. Ding and Harini Suresh. 2025. The malicious technical ecosystem: Exposing limitations in technical governance of AI-generated non-consensual intimate images of adults. In \textit{CHI'25 Sociotechnical AI Governance Workshop (STAIG ’25)}, April 2025, Yokohama, Japan.
\end{minipage}
\end{center}
\section{Introduction}\label{intro}
AI-Generated Non-Consensual Intimate Images (AIG-NCII), colloquially known as ``deep fake pornography,'' are a form of image-based sexual abuse that disproportionately harm and silence women, girls, LGBTQ+ people, and racial minorities \cite{flynn_deepfakes_2022, center_for_democracy_and_technology_ibsa_2025,santiago_lakatos_revealing_2023, mcglynn_beyond_2017}. Image-based sexual abuse covers a spectrum of behaviors including the non-consensual \textit{creation} of intimate images, non-consensual \textit{distribution} of intimate images, and \textit{threatening} to distribute those images \cite{mcglynn_beyond_2017,qin_did_2024}. Recognizing the extensive prior work on preventing AI-generated child sexual abuse materials (AIG-CSAM) and non-consensual \textit{distribution} of intimate images, this paper aims to contribute to the unique challenge of preventing the \textit{creation} of AIG-NCII of \textit{adults} \cite{thiel_identifying_2024,qin_did_2024,akerley_lets_2021,becca_branum_ndii_2024,thorn_all_tech_is_human_safety_2024}. These include the documented 35,545 images depicting 26 members of congress and thousands more depicting celebrities, public figures, and ordinary women and teens \cite{thiel_identifying_2024,thorn__weprotect_global_alliance_evolving_2024,thorn_all_tech_is_human_safety_2024,american_sunlight_project_deepfake_2024,henry_ajder_state_2019,genevieveoh_dataset_2023}. While recent advancements in corporate image-generation models like Stability AI's Stable Diffusion have resulted in alarmingly photorealistic AIG-NCII, there has existed since 2017 a prolific ecosystem of open-source face-swapping models such as DeepFaceLab, DeepNude, and FaceSwap that support nearly 200 ``nudifying'' software programs allowing non-technical users to create what they call ``deep fake pornography'' within minutes \cite{thiel_identifying_2024,perov_deepfacelab_2020,genevieveoh_dataset_2023,torzdf_faceswap_2017,iperov_deepfacelab_2018,yuanxiao_deepnude_2020,mehta_can_2023,timmerman_studying_2023}. This decentralized technical ecosystem of open-source models and tools (which we characterize as the ``malicious technical ecosystem'' or ``MTE'') enables the creation of \textit{AIG-NCII of adults} that are often noticeably ``fake'' yet \textit{still a form of abuse} that results in mental and physical harm, reputational damage, financial costs, and disruption of social relationships \cite{santiago_lakatos_revealing_2023,sophie_cockerham_deepfake_2022,mcglynn_its_2021,umbach_prevalence_2025,akerley_lets_2021,flynn_deepfakes_2022}. Image-based sexual abuse, like other forms of technology-facilitated gender-based violence, also causes a gendered chilling effect where victim-survivors retreat from online spaces \cite{sambasivan_they_2019,mcglynn_beyond_2017,kristine_baekgaard_technology-facilitated_2024}. These effects already appear in the case of AIG-NCII as seen from interviews and testimonials \cite{flynn_deepfakes_2022,genevieveoh_dataset_2023}. 

Despite extensive documentation of the harms and scale of these models and tools, they remain largely ungoverned by current technical prevention methods. In this paper, we locate the MTE as a unique challenge for the sociotechnical governance of adult AIG-NCII. First, we identify and contextualize the MTE within the broader synthetic content pipeline of creation, distribution, and consumption. Next, by drawing on the National Institute of Standards and Technology (NIST) AI 100-4 report on governing synthetic content, we show how current governance mechanisms fail to effectively regulate the MTE: 1) \textit{broad synthetic content governance measures} focus on transparency, which is insufficient to address the harms of watermarked, human-detectable adult AIG-NCII, 2) \textit{AIG-NCII specific governance measures} conflate CSAM and adult NCII, proposing methods more effective in regulating CSAM than adult NCII, and 3) \textit{adult-specific AIG-NCII governance measures} acknowledge the existence of the MTE but propose methods that will primarily regulate large corporate models. 
Overall, we stress the need for governance frameworks that cite AIG-NCII of adults as a high-risk harm to assess their effectiveness in regulating the MTE by considering the three listed limitations.
\begin{figure*}[t]
  \centering
  \includegraphics[width=\textwidth]{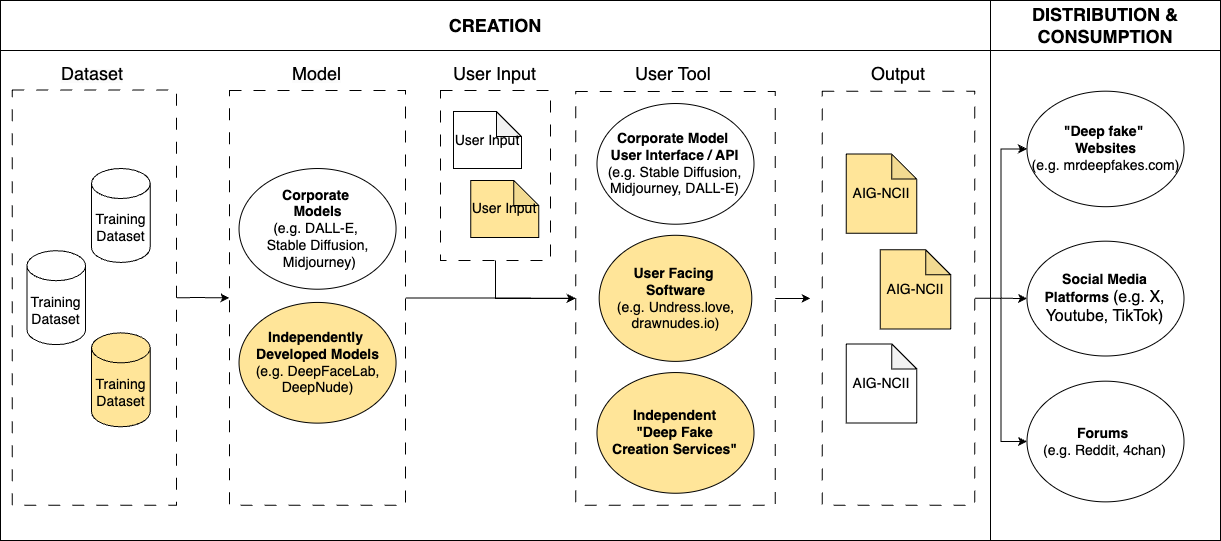}
  \caption{Supply chain for AI Generated Non-Consensual Intimate Images that expands on the NIST AI 100-4 synthetic content pipeline of creation, distribution, and consumption. Yellow circles represent the malicious technical ecosystem (MTE). Training dataset, user input, and AIG-NCII are highlighted to represent artifacts within the pipeline and not any specific example.}
  \Description{Supply chain for AI Generated Non-Consensual Intimate Images (AIG-NCII) that expands on the NIST AI 100-4 synthetic content pipeline (creation, distribution, and consumption) showing the key technical components involved in each stage from dataset, model, user input, user tool, to output in creation and then various websites, platforms, and forums for distribution and consumption. Yellow highlighted circles represent the core of the malicious technical ecosystem (MTE). Note: while ``deep fake websites'' in distribution/consumption are closely associated with the MTE as the primarily destination of AIG-NCII, we do not include them within the MTE because their governance requires downstream response, not technical prevention.}
  \label{fig:ecosystem}
\end{figure*}
\section{Response vs. Prevention: Locating the Role of Sociotechnical AI Governance in Mitigating AIG-NCII}\label{responseprevention}
Before diving into the challenges of technical prevention, it is important to contextualize the role of sociotechnical AI governance within the larger multi-stakeholder space of AIG-NCII mitigation efforts. The issue of AIG-NCII, like other forms of image-based sexual abuse and technology-facilitated gender-based violence, requires both \textit{prevention} and \textit{response} efforts. There are already many powerful \textit{responses} to AIG-NCII, including: two federal legislation efforts that both passed the Senate (the DEFIANCE Act and the Take It Down Act) which require expedited platform take-down procedures and create civil liability for malicious actors; advocacy petitions for the take-down of "deepfake pornography" apps and sites; and voluntary platform trust and safety commitments \cite{sen_cruz_ted_r-tx_take_2023,sen_durbin_richard_j_d-il_defiance_2023,my_image_my_choice_blockmrdeepfakes_nodate,partnership_on_ai_responsible_2023,center_for_democracy_and_technology_ibsa_2025}. While these response measures are critical to addressing the downstream distribution and consumption of AIG-NCII, they place the burden of removal on survivors, many of whom may be unwilling to report abuse, hesitant to engage with the criminal justice system, and/or lack the emotional, financial, and physical capacity to engage in lengthy take-down procedures or legal action \cite{flynn_deepfakes_2022,lorenz_qualitative_2019}. Public attitudes toward ``deepfake pornography'' often involve victim-blaming and stigmatization which also result in less help-seeking by victim-survivors \cite{umbach_prevalence_2025}. Most importantly, responsive measures that only target the distribution fail to govern the technologies that \textit{create} AIG-NCII, thus allowing an unimpeded stream of new AIG-NCII for survivors to take down. \textit{Preventing the creation} of AIG-NCII, then, is a key a challenge for sociotechnical AI governance.
\section{The Malicious Technical Ecosystem (MTE)}\label{MTE}
This section will characterize the MTE of adult AIG-NCII and contextualize it within a broader synthetic content pipeline. Figure \ref{fig:ecosystem} maps the AIG-NCII supply chain onto NIST's synthetic content pipeline of \textit{creation, distribution,} and \textit{consumption} \cite{nist_reducing_2024}. As our focus is on technical prevention, we add additional granularity to \textit{creation} by including specific artifacts such as training datasets, models, and user tools. The highlighted circles represent the core of the MTE: independently developed models, user-facing software, and independent "deep fake" creation services.

\subsection{Independently Developed Models}\label{models}
A current quick search of ``deepfake'' on GitHub returns 7.2k repositories, a majority identified to generate NSFW content \cite{patrick_trueman_esq_github_2023}. The most popular NCII generator is DeepFaceLab, a GAN-based face-swap model with 17.2k stars on GitHub \cite{iperov_deepfacelab_2018,timmerman_studying_2023}. The creators of DeepFaceLab also built MrDeepFakes.com to encourage the use of their codebase \cite{sophie_cockerham_deepfake_2022}. This site is currently the largest dedicated ``deepfake pornography'' website with over 1 billion accumulated views from 2016-2023 \cite{genevieveoh_dataset_2023}. Other codebases that use similar architectures include FaceSwap and DeepNude \cite{torzdf_faceswap_2017,yuanxiao_deepnude_2020}. While some repositories like FaceSwap includes a ``zero-tolerance'' policy for NSFW content, there are no real technical barriers preventing AIG-NCII generation \cite{torzdf_faceswap_2017}. These models have trained on various datasets including Flickr-Faces-HQ, nsfw-data-scraper, and nsfw-data-source-urls \cite{tero_karras_flickr-faces-hq_2019,alex_kim_nsfw_data_scraper_2019,evgeny_bazarov_nsfw_data_source_urls_2019}. A study of DeepNude also found that the algorithm ``cannot perform similar translations on images of men, having been specifically trained on images of women'' \cite{henry_ajder_state_2019}.
\subsection{User Facing Software and Independent ``Deep Fake Creation Services''}\label{tools}
Building upon these open-source independently developed models are over 200 dedicated AIG-NCII generation tools, or ``nudifier services,'' that enable non-technical audiences to create AIG-NCII of any user-uploaded image within minutes \cite{genevieveoh_dataset_2023,sophie_cockerham_deepfake_2022,mehta_can_2023}. There are also over 1700 third party paid ``Deep Fake Creation Services,'' making up a ``fully fledged online industry'' \cite{genevieveoh_dataset_2023,santiago_lakatos_revealing_2023}. The outputted AIG-NCII, often watermarked, labeled ``AI-generated'', or captioned ``not [\textit{depicted person}],'' are then distributed and consumed across various channels, including 50+ dedicated "deep fake pornography" websites, social media platforms like X, YouTube, and TikTok, and discussion forums like Reddit and 4chan \cite{henry_ajder_state_2019,genevieveoh_dataset_2023}. 
\subsection{Supporting Technological Infrastructure}
This paper is focused on technical prevention in the form of governing \textit{models and tools}. However, the MTE is also supported by infrastructure like Github to host the codebases and chat forums (Reddit, 4chan, 8chan, and Voat) to guide new developers \cite{patrick_trueman_esq_github_2023,cinecomnet_deepfake_2019,henry_ajder_state_2019,timmerman_studying_2023}. Google  consistently lists "deepfake creation" tools and services the at top of search results \cite{kat_tenbarge_found_2023, genevieveoh_dataset_2023}. Mastercard and Visa enable malicious actors to monetize their creation services \cite{kat_tenbarge_found_2023}. Civil society organizations are also calling for these tech companies to take accountability for their support of the ``MTE'' \cite{patrick_trueman_esq_github_2023,genevieveoh_dataset_2023}.
\section{Implications of the MTE: Limitations in Current Synthetic Content Governance Practices}\label{governance}
In this section, we anticipate three ways in which current governance practices, as reflected in the 2024 NIST report on ``Reducing Risks Posed by Synthetic Content,'' may fail to regulate the MTE \cite{nist_reducing_2024}. NIST begins with broadly applicable practices for digital content transparency for synthetic content, along with a specific section on technical prevention methods for AIG-NCII and AIG-CSAM that includes: 1) training data filtering, 2) input data filtering, 3) output filtering, 4) hashing confirmed AIG-CSAM and AIG-NCII, and 5) provenance data tracking techniques.

\subsection{Limitation 1: Synthetic content governance measures focus on transparency, which is insufficient to address the harms of human-detectable and overtly watermarked adult AIG-NCII.}\label{lim1} 
Models within the MTE like DeepFaceLab and DeepNude are not powered by state-of-the-art image-generation models trained on massive datasets. For comparison, Stable Diffusion's training dataset LAION-5B is 28,000 times larger than curated image datasets like Flickr-Faces-HQ used by DeepFaceLab \cite{schuhmann_laion-5b_2022,tero_karras_flickr-faces-hq_2019,perov_deepfacelab_2020,thiel_identifying_2024}. As a result, the images produced within the MTE can often be detected as ``fake'' by a viewer. Many tools within the MTE further watermark and label AIG-NCII outputs as ``AI-generated'' or ``FAKE'' \cite{yuanxiao_deepnude_2020}. Transparency practices like provenance data tracking, synthetic content detection, and user education labels are suitable for governing other synthetic content, such as political deep fakes and voice cloning, where harms like disinformation, election security risks or fraudulent transactions are dependent on the detectability of outputs. These efforts may also aid in accountability and content moderation during distribution and consumption of AIG-NCII. 
However, AIG-NCII governance measures that stop at transparency make the flawed assumption that noticeably ``fake'' content is no longer harmful enough to regulate. This sentiment also belongs to deepfake creators themselves. One said, ``I can see how some women would have psychological harm from this, but they can just say, `It’s not me, this has been faked, I can’t suffer any damages from this''' \cite{sophie_cockerham_deepfake_2022}. On the contrary, ``fake'' NCII can still result in mental and physical harm, reputational damage, financial costs, and further cyber-violence \cite{flynn_deepfakes_2022,genevieveoh_dataset_2023}. Most critically, ``deepfake pornography,'' like other forms of technology-facilitated gender based violence, creates a gendered chilling effect that disproportionately silences gender, racial, and sexual minorities \cite{maddocks_deepfake_2020}. A study on public attitudes further found that ``labeling pornographic deepfakes as fictional did not mitigate the videos’ perceived wrongfulness'' \cite{matthew_b_kugler_deepfake_2021}.

\subsection{Limitation 2: AIG-NCII specific measures conflate CSAM and adult NCII, proposing methods more effective in regulating CSAM than adult NCII.}\label{lim2}
Adult AIG-NCII and CSAM are often conflated into one category by NIST and other multistakeholder frameworks \cite{partnership_on_ai_responsible_2023,center_for_democracy_and_technology_ibsa_2025}. However, examples of effective prevention methods primarily target CSAM. For example, the key study cited in "Method 1: Training Data Filtering" uses Microsoft’s PhotoDNA tool to query existing databases of known illegal CSAM images \cite{microsoft_photodna_2015,thiel_identifying_2024}. With CSAM, consent is not debatable: any sexually explicit depiction of a minor is illegal under federal child pornography law \cite{us_department_of_justice_citizens_2023}. Thus, AIG-CSAM tools can draw upon existing law enforcement databases of CSAM. With adults, it is more difficult to collect such a database as consent is not automatically reflected in the dataset. Although CSAM and adult-NCII methods may be similar, it is important for any governance framework or method that aims to be applicable to \textit{both} CSAM and adult-NCII to consider this distinction. 

\subsection{Limitation 3: Adult-specific AIG-NCII governance measures acknowledge the existence of the MTE but propose methods that will only regulate large corporate models.}\label{lim3}
While NIST also references the same models and tools described in our MTE, their governance methods for adult NCII primarily apply to the architecture of larger corporate image-generation models. For example, ``Input Data Filtering'' methods like keyword blocking and ``Red-Teaming and Testing'' methods like adversarial prompt generation assume a prompt-based image-generation model. Therefore, these methods are inapplicable for the MTE's ``deep fake pornography creation tools'' where user input is just the photo they want to ``nudify.'' The gap in governance here is primarily caused by the assumption that the image-generation technology is trustworthy and NCII is a result of malicious users. In the MTE, the \textit{technology itself} is malicious, developed intentionally for a single goal of generating NCII. Frameworks or methods that aim to prevent AIG-NCII should consider their applicability to the ``MTE'' \textit{in addition} to corporate models. 

\section{Looking Forward: Towards Survivor-Centered Prevention of Adult AIG-NCII}\label{conclusion}
Survivors and advocates have and continue to fight long and difficult battles against image-based sexual abuse and technology-facilitated gender-based violence. This paper builds upon their tremendous work by calling upon the sociotechnical research community to prioritize the improvement of current governance methods that fall short in effectively regulating the malicious technical ecosystem that produces tens of thousands of life-threatening adult AIG-NCII reaching billions of viewers every year. Under the MTE paradigm, a harmful image can be reported and removed  or a particular ``deepfake'' repository can be taken down, but just as quickly, new ones can spring up. In this paper, we point out three ways in which current synthetic content governance methods fail to govern the MTE producing adult AIG-NCII content. More importantly, we expose some \textit{flawed fundamental assumptions} behind these governance methods. The MTE and its terrible consequences show us that fake content is \textit{still} harmful content worthy of technical governance, and that harmful outputs are not just products of malicious users, but of malicious technologies themselves. 
\bibliographystyle{ACM-Reference-Format}
\bibliography{body} 
\end{document}